# Approach combining the Rietveld method and pairs distribution function analysis to study crystalline materials under high-pressure and/or temperature: Application to rhombohedral $Bi_2Te_3$ phase

J. C. de Lima[1,a)], Z. V. Borges[1], C. M. Poffo[2], S. M. Souza[3], D. M. Trichês[3], and R.S. de Biasi[4]

[1]*Departamento de Engenharia Mecânica, Universidade Federal de Santa Catarina, Programa de Pós-Graduação em Ciência e Engenharia de Materiais, Campus Universitário Trindade, C.P. 476, 88040-900 Florianópolis, Santa Catarina, Brazil*

[2]*Universidade Federal de Santa Catarina, Campus de Araranguá, 88900-000 Araranguá, Santa Catarina, Brazil.*

[3]*Departamento de Física, Instituto de Ciências Exatas, Universidade Federal do Amazonas, 69077-000 Manaus, Amazonas, Brazil*

[4]*Seção de Engenharia Mecânica e de Materiais, Instituto Militar de Engenharia, 22290-270 Rio de Janeiro, RJ, Brazil*

ABSTRACT

An approach combining the Rietveld method and pairs distribution function analysis to study crystalline materials under high-pressure and/or temperature was early proposed by us, and in this study, it was applied to investigate de effect of high-pressure on the rhombohedral $Bi_2Te_3$ phase. The refined structural parameters obtained from the Rietveld refinement of the XRD patterns measured for pressures up to 9.1 GPa were used as input data to simulate the partial and total structure factors $S_{Bi-Bi}(K)$, $S_{Bi-Te}(K)$, $S_{Te-Te}(K)$, and $S_{Bi2Te3}(K)$. Fourier transformation of the $S_{ij}(K)$ factors permitted to obtain the partial and total pairs distribution functions $G_{Bi-Bi}(R)$, $G_{Bi-Te}(R)$, $G_{Te-Te}(R)$, and $G_{Bi2Te3}(R)$. The first coordination shells of these $G_{ij}(R)$ functions are formed by subshells and, with increasing pressure in the 1.1−6.3 GPa range, occur a partial separation of subshells. Also, the increase of pressure in this range promotes a drastic reduction in the values of the intralayer angles Te-Bi-Te, and consequently, in the intralayer distance Te-Te. A drastic reduction in the interlayers distance Te-Te was also observed. Several studies are reported in the literature, including one carried out by us, show the presence of an ETT in this pressure range. The obtained results suggest that the ETT is related with the decrease of the intralayer angles Te-Bi-Te, and intra- and interlayer distance Te-Te.

Experimental results describing the pressure dependence up to ≈ 6 GPa of the thermoelectric power (S), electrical resistivity (ρ), and power fator (PF) for pure rhombohedral $Bi_2Te_3$ and $(Bi,Sb)_2(Te,Se)_3$ alloys are reported in the literature, and an enhancement of the power factor in the 1.1−6.3 GPa range is observed. The results obtained in this study give evidence that this enhancement in the power factor is related with the decrease of the intralayer angles Te-Bi-Te, and with the decrease of intralayer- and interlayers homopolar Te-Te bonds.

a) Author to whom correspondence should be addressed. Electronic mail: joao.cardoso.lima@ufsc.br or jcardoso.delima@gmail.com

# I. INTRODUCTION

Despite the extraordinary progress in the development of DACs, allowing that several dozen GPa be achieved, they have yet a limited window of few tens (two or three) of degrees, imposing a restriction in the angular range (2$\theta$) in which the X-ray diffraction (XRD) patterns are measured. This limitation makes difficult to visualize the effect of high-pressure on the structure being studied, as well as it increases the difficulties in the determination of new high-pressure structure phases emerging due to the few number of diffraction peaks recorded. Trying to overcome the difficulties mentioned above, at least partially, we implemented an approach combining the Rietveld method (RM) [1] and the pairs distribution function (PDF) analysis [2−5]. This approach is well detailed in Ref. [1,6,7], and it will not be repeated here.

The layered isostructural rhombohedral $Bi_2Te_3$, $Sb_2Te_3$, $Bi_2Se_3$ compounds and alloys based on them have great potential for applications in thermoelectric (TE) devices like solid-state coolers or generators at temperatures near the ambient [8].

We produced the TE isostructural nanostructured $Bi_2Te_3$, $Sb_2Te_3$, and $Bi_2Se_3$ compounds via mechanical alloying technique, and the effect of high-pressure on them was investigated. The reached results for $Sb_2Te_3$ were reported in Ref. [9]. For $Bi_2Te_3$, despite new high-pressure phases have been observed, only the effect of high-pressure up to 10 GPa on the rhombohedral structure was reported [10]. By considering the Birch-Murnaghan's equation of state (BM-EOS) linearization versus the Eulerian strain, an electronic topological transition (ETT) was observed at ≈ 3.2 GPa. An ETT at ≈ 3.7 GPa was also observed for $Sb_2Te_3$. High-pressure study on $Bi_2Se_3$ will be carried out soon. Recently, Ovsyannikov *et al*. [11] reported experimental results for pure rhombohedral $Bi_2Te_3$ and $(Bi,Sb)_2(Te,Se)_3$ alloys. Those researchers showed an enhancement of power factor for pressures smaller than 6 GPa.

The approach combining the Rietveld method and pairs distribution function analysis permitted to build the partial pairs distribution functions (PPDFs) as well as the total pairs distribution function (PDF). Details of these functions are described in Ref. [1]. We believe that the PPDFs and PDF functions can give a more accurate insight view of the effect of high-pressure on the structure. Thus, the main goal of the present study is to apply the approach combining the Rietveld method and pairs distribution function analysis to rhombohedral $Bi_2Te_3$ phase in order to understand the effect of high-pressure on this structure. In addition, it will be tried to associate, at least partially, the structural changes promoted by the application of high-pressure on the structure with the changes observed in the macroscopic thermoelectric power, electrical resistivity, and power factor properties reported in Ref. [11].

## II. STRUCTURAL DATA FOR THE RHOMBOHEDRAL $Bi_2Te_3$ PHASE

At room temperature and atmospheric pressure, the $Bi_2Te_3$ compound crystallizes in a trigonal/rhombohedral structure (space group R-3m H (166)), with the Bi atoms occupying the Wyckoff site *6c* (0,0,Z) (in the hexagonal setting) and Te atoms occupying the *3a* (0,0,0) and *6c* (0,0,Z) sites [12]. In Fig. 1 of Ref. [9], it is shown the conventional unit cell for $Sb_2Te_3$. The c-hexagonal axis is perpendicular to the plane of the layers. Within each sheet the chemical bonds are of the ionocovalent type, while those between the sheets (inter-sheets) and between the layers (interlayers) are of the van der Waals type [9,10]. In previous paper [10], we studied the effect of high-pressure on rhombohedral $Bi_2Te_3$ for pressures up to 10 GPa, and considering the Birch-Murnaghan's equation of state (BM-EOS) linearization versus the Eulerian strain, an electronic topological transition (ETT) was observed at ≈ 3.2 GPa. Table 1 lists the values of lattice parameters and Z-coordinates of Wickoff position *6c* (0,0,Z) for the Bi and Te atoms as a function of pressure applied.

## III. EXPERIMENTAL PROCEDURE AND SIMULATIONS

For $Bi_2Te_3$, the structural data as a function of pressure are same used in Ref. [10]. All the details about the sample preparation and XRD measurements are described in this reference, and therefore, they will not be repeated here.

The shell structures for the rhombohedral $Bi_2Te_3$ phase for each value of pressure applied were calculated up to *Rmax* = 25 Å (arbitrarily chosen) using the crystallographic data space group (R-3m H (166), the lattice parameters *a* and *c* listed in Table 1, and the Wickoff positions Bi *6c* (0,0,Z), Te *6c* (0,0,Z), Te *3a* (0,0,0) as input data in the Crystal Office software® [13], and using the software′s tools "*Output + Coordinates + Shell Structure*". By putting the position Bi *6c* (0,0,Z) at origin, the Bi-Bi and Bi-Te coordination numbers and interatomic distances were calculated; for Te *6c* (0,0,Z) and Te *3a* (0,0,0) at origin separately, the Te-Te and Te-Bi coordination numbers and interatomic distances were calculated. To simulate the partial structure factors $S_{ij}(K)$, the calculated coordination numbers and interatomic distances Bi-Bi, Bi-Te, and Te-Te were used as input data in a computational FORTRAN code describing the expression (1) of Ref. [1]. Based on this reference, a value *Kmax* = 30 Å$^{-1}$ was chosen. Due to the fact that $S_{Bi-Te}(K) = S_{Te-Bi}(K)$, in this study, the $S_{Bi-Te}(K)$ and $S_{Te-Te}(K)$ factors were averaged. By Fourier transformation of the simulated $S_{ij}(K)$ factors, the partial and total reduced total distribution functions $\gamma_{ij}(R)$ and $\gamma(R)$, the partial and total pairs distribution functions $G_{ij}(R)$ and $G(R)$, and partial and total radial distribution functions $RDF_{ij}(R)$ and $RDF(R)$ were obtained. To obtain the $S_{Bi2Te3}(K)$, the $W_{Bi-Bi}(K)$, $W_{Bi-Te}(K)$, and $W_{Te-Te}(K)$ weights were calculated. All the computational FORTRAN codes describing the expressions given in Section II of Ref. [1] were written by one of the authors (J.C. de Lima).

## IV. RESULTS AND DISCUSSION

Figures 1-4 shows the $G_{Bi-Bi}(R)$, $G_{Bi-Te}(R)$, $G_{Te-Te}(R)$, and $G_{Bi2Te3}(R)$ functions with increasing pressure up to 9.1 GPa. From the calculated shell structures, at ambient pressure (0 GPa), the $G_{Bi-Bi}(R)$ function (see Fig. 1) shows that the first coordination shell is formed by two subshells partially overlapped containing 6 and 3 Bi-Bi pairs at 4.390 Å and 4.763 Å, respectively; the second coordination shell is formed by two subshells partially overlapped containing 1 and 3 Bi-Bi pairs at 6.121 Å and 6.477 Å, respectively, and the third coordination shell is formed by three subshells partially overlapped containing 6 Bi-Bi pairs at 7.532 Å, at 7.604 Å, and at 7.825 Å. It is interesting to note that, for pressures between 1.1 and 6.3 GPa, the subshells forming the first and second coordination shells are well separated, while in the third coordination shell occurs a tendence of subshells to separate.

From the calculated shell structures, at ambient pressure, the calculated $G_{Bi-Te}(R)$ function (see Fig. 2) shows that the first coordination shell is formed by two subshells overlapped containing each one 3 Bi-Te pairs at 3.050 Å and at 3.239 Å; the second coordination shell is formed by four subshells partially overlapped containing 3 Bi-Te pairs at 5.097 Å, at 5.346 Å, 5.456 Å, and 1 Bi-Te pair at 5.731 Å; and the third coordination shell is formed by four subshells partially overlapped containing 3 Bi-Te pairs at 6.727 Å, 6 Bi-Te pairs at 6.917 Å, at 7.003 Å, and at 7.219 Å. For pressures between 1.1 and 6.3 GPa, in the three coordination shells occur a tendence of subshells to separate.

From the calculated shell structures, at ambient pressure, the calculated $G_{Te-Te}(R)$ function (see Fig. 3) shows that the first coordination shell is formed by two subshells overlapped containing each one 6 Te-Te pairs at 4.390 Å and at 4.496 Å; the second coordination shell is formed by two subshells separated containing 6 Te-Te pairs at 6.284 Å and 2 Te-Te pairs at 6.439 Å; and the third coordination shell is formed by two subshells separated containing 7 Te-Te pairs at 7.666 Å, and 12 Te-Te pairs at 7.793 Å. For pressures between 1.1 and 6.3 GPa, in the first and second coordination shells occur a tendence of subshells to separate, while in the third coordination shell occur a tendency of subshells to overlap. It is interesting to note that the ETT reported in Ref. [10] is observed in this pressure range (at about P ≈ 3.2 GPa).

The calculated $G_{Bi2Te3}(R)$ function as a function of pressure applied is shown in Fig. 4. For pressures between 1.1 and 6.3 GPa, the first coordination shell is splitted into two subshells partially overlapped; the second coordination shell becomes clear that it is formed by three subshells well separated and intense; and the third coordination shell becomes clear that it is formed by, at least, five subshells partially overlapped.

Figure 1 of Ref. [9] shows the conventional unit cell for $Sb_2Te_3$, which is similar to the $Bi_2Te_3$ one. Figure 5 (a)-(b) shows the pressure dependence of the smallest angle Te(S1)-Bi(S2)-Te(S3) in the layer 2 (L2) and the interatomic distances Te(S1)-Bi(S2), Bi(S2)-Te(S3) and Te(S1)-Te(S3), respectively. The inset of Fig. 5(a) shows the angle and sheets (S) of L2. From the Fig. 5(a) one can see that, for pressures between 0.1 and 3 GPa, the values of angle decreases, and for pressures up to 6.3 GPa, it increases. For pressures larger than 6.3 GPa, the values oscillate. Figure 5(b) shows that the interatomic distance Te(S1)-Bi(S2) is more sensitive to the effect of high-pressure than the Bi(S2)-Te(S3) one. The decrease in the interatomic distance Te(S1)-Te(S3) for pressures between 0.1 GPa and 4.6 GPa refletcs the decrease in the angle Te(S1)-Bi(S2)-Te(S3), and consequently, suggest that the electronic repulsion begins to play an important role. The pressure dependence of the smallest angle Te(S3)-Bi(S4)-Te(S5) in L2 and of interatomic distances Te(S3)-Bi(S4), Bi(S4)-Te(S5) and Te(S3)-Te(S5) has the same behavior. However, the value of this angle is slightly smaller, as shown in Fig. 6(a)-(b). Figure 7 shows the pressure dependence of the interlayers interatomic distance Te(L2-S5)-Te(L3-S5). One can see that, for pressures between 0.1 GPa and 2.5 GPa, it reaches a maximum; it decreases fastly up to 4.6 GPa, and oscillating slowly for larger pressures. The decrease shows that the electronic repulsion plays an important role. Again, it is interesting to note that the ETT reported in Ref. [10] is observed in this pressure range (at about P ≈ 3.2 GPa).

It is interesting note that the decrease of the intralayer angles Te-Bi-Te, and consequently, the decrease of the intralayer homopolar Te-Te bonds (see Figs. 5 and 6) promote the splitting of the first coordination shell of $G_{Te-Te}(R)$ function into two subshells partially overlapped that, at ambient pressure, is formed 6 Te-Te pairs at 4.390 Å and 6 Te-Te pairs at 4.496 Å (see Fig. 3). The same effect is observed in the first coordination shell of $G_{Bi-Te}(R)$ function, which is splitted into two subshells partially overlapped that, at ambient pressure, is formed by 3 Bi-Te pairs at 3.050 Å and 3 Bi-Te pairs at 3.239 Å (see Fig. 2). The effect in $G_{Bi-Bi}(R)$ function is to distance the two subshells that, at ambient pressure, contain 6 Bi-Bi pairs at 4.390 Å and 3 Bi-Bi pairs at 4.763 Å (see Fig. 1). These results indicate that the repulsive part of the intralayer and interlayers homopolar Te-Te bonds plays an important role in the structural stability of the rhombohedral $Bi_2Te_3$ phase.

It is interesting to try to correlate the effect of high pressures previously shown with macroscopic thermoelectric properties of $Bi_2Te_3$. Recently, Ovsyannikov *et al.* [11] reported experimental results describing the pressure dependence of the thermoelectric power (S), electrical resistivity (ρ), and power fator (PF) for pure rhombohedral $Bi_2Te_3$ and $(Bi,Sb)_2(Te,Se)_3$ alloys, which are shown in Fig. 8(a)-(e). The inset in figure (d) is amplified on the lower right side. For this study, these properties will be only considered for pure rhombohedral $Bi_2Te_3$ and for pressures up to 10 GPa. From the point of view of technological applications, the thermoelectric materials are characterized by the dimensionless figure of merit ZT = $(S^2\sigma)T/k$, where $\sigma$, k and T are the electrical conductivity, the termal conductivity and the absolute temperature, respectively. The $S^2\sigma$ is named power fator (PF). For thermoelectric applications, ZT > 1 is desired. From the Fig. 8(a), curve #2, one can see that the thermoelectric power S decreases with increasing pressure up to ≈ 6.5 GPa, but positive bumps are observed for pressures P ≈ 1 GPa and 3 < P < 6.5 GPa. From the Fig. 8(d), curve #2, and of amplified inset one can see that the electrical resistivity decreases with increasing pressure, while $\sigma = 1/\rho$ increases for the pressure range shown. From the Fig. 8(e), curve #2, one can see that the PF curve shows positive bumps for pressures 1 < P < 2 GPa and 3 < P < 5 GPa. Figures 5(a) and 6(a) show that, for pressures 0.1 < P < 4 GPa, both angles L2 [Te(S1)-Bi(S2)-Te(S3)] and L2 [Te(S3)-Bi(S4)-Te(S5)] decrease drastically. The drastic reduction in these angles is corroborated by the reduction in the homopolar interatomic distances Te(S1)-Te(S3) and Te(S3)-Te(S5) in this pressure range, as shown in Figs. 5(b) and 6(b). From the Fig. 7, one can see that the interlayers interatomic distance Te(L2-S5)-Te(L3-S5) decreases for pressures 1 < P < 4 GPa.

The results obtained in this study for the rhombohedral $Bi_2Te_3$ phase, for 1 < P < 4 GPa, give strong evidence that the ETT reported in our previous study [10] at ≈ 3.2 GPa and also the enhancement of the power factor for pressures smaller than 6 GPa reported in Ref [11] are related with the decrease of the intralayer angles Te-Bi-Te, and with the decrease of intralayer- and interlayers homopolar Te-Te bonds.

## V. CONCLUSIONS

This study reports the results obtained of application of an approach combining the Rietveld method and pairs distribution function analysis to study crystalline materials under high-pressure and/or temperature to rhombohedral $Bi_2Te_3$. The refined structural parameters obtained from the Rietveld refinement of the XRD patterns measured for pressures up to 9.1

GPa were used as input data to simulate the partial and total structure factors $S_{Bi-Bi}(K)$, $S_{Bi-Te}(K)$, $S_{Te-Te}(K)$, and $S_{Bi2Te3}(K)$. Fourier transformation of them permitted to obtain the partial and total pairs distribution functions $G_{Bi-Bi}(R)$, $G_{Bi-Te}(R)$, $G_{Te-Te}(R)$, and $G_{Bi2Te3}(R)$. The first coordination shells of these $G_{ij}(R)$ functions are formed by subshells and, with increasing pressure in the 1.1−6.3 GPa range, occur a partial separation of subshells. Also, the increase of pressure in this range promotes a drastic reduction in the values of the intralayer angles Te-Bi-Te, and consequently, in the intralayer distance Te-Te. Also, a drastic reduction in the interlayers distance Te-Te is observed. Several studies reported in the literature, including one carried out by us, show the presence of an ETT in this pressure range. The results obtained in this study suggest that the ETT is related with the decrease of the intralayer angles Te-Bi-Te,and intra- and interlayer distance Te-Te.

Experimental results describing the pressure dependence up to ≈ 6 GPa of the thermoelectric power (S), electrical resistivity (ρ), and power fator (PF) for pure rhombohedral $Bi_2Te_3$ and $(Bi,Sb)_2(Te,Se)_3$ alloys are reported in the literature. In the 1.1−6.3 GPa range, an enhancement of the power factor is observed. The results obtained in this study give strong evidence that this enhancement of the power factor is related with the drastic decrease of the intralayer angles Te-Bi-Te, and with the decrease of the intra- and interlayers homopolar Te-Te bonds.

**ACKNOWLEDGMENTS**

One of the authors (Z. V. Borges) was financially supported by a scholarship from CNPq.

TABLE

Table 1: Structural data as a function of pressure for rhombohedral $Bi_2Te_3$.

| Pressure (GPa) | Lattice Parameters | | 6c (0, 0, Z) | |
| --- | --- | --- | --- | --- |
| | $a$(Å) | $c$(Å) | Bi | Te |
| 0 | 4.39 | 30.46 | 0.39955 | 0.21138 |
| 0.1 | 4.3815 | 30.3611 | 0.40039 | 0.20987 |
| 1.1 | 4.3466 | 30.0697 | 0.40162 | 0.21105 |
| 2.5 | 4.3028 | 29.7538 | 0.40191 | 0.21098 |
| 4.6 | 4.247 | 29.4149 | 0.40338 | 0.20763 |
| 6.3 | 4.2092 | 29.2509 | 0.40334 | 0.20777 |
| 7.7 | 4.1765 | 29.1259 | 0.40026 | 0.20849 |
| 9.1 | 4.1495 | 29.1024 | 0.40167 | 0.20892 |

FIGURES

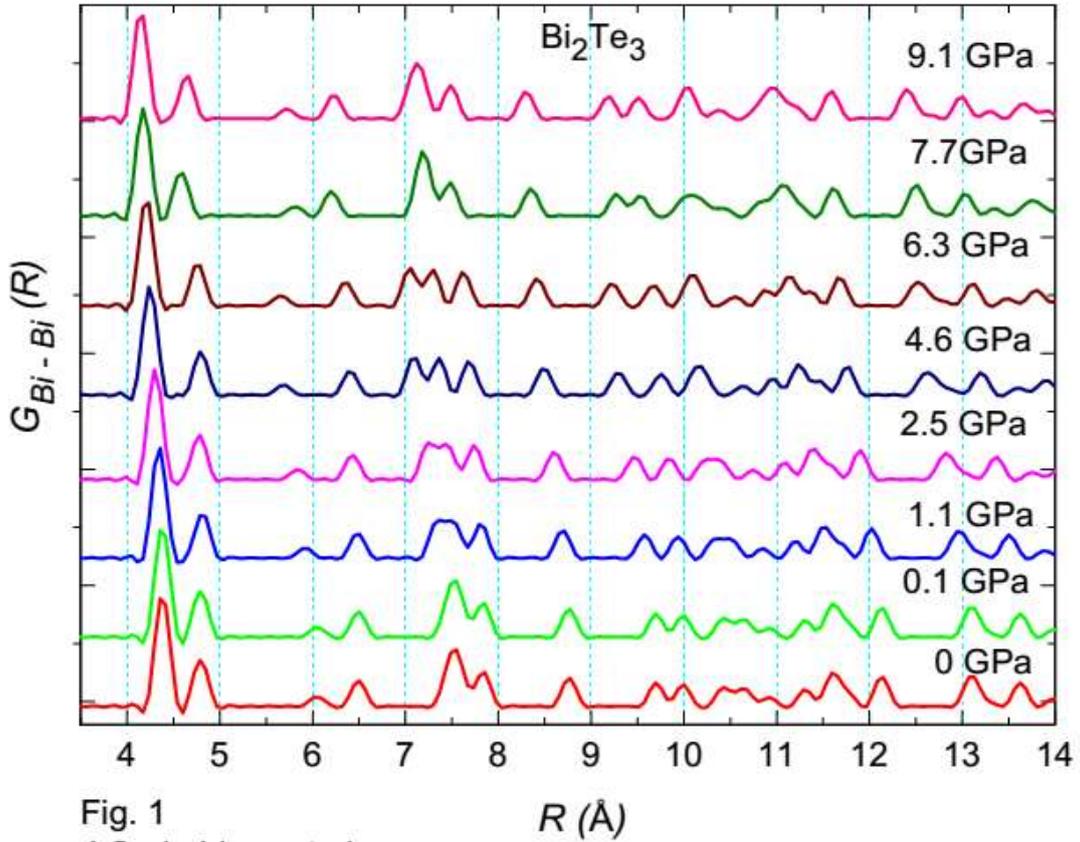

Fig. 1
J.C. de Lima *et al.*

Figure 1: Pressure dependence of $G_{Bi-Bi}(R)$ function for rhombohedral $Bi_2Te_3$.

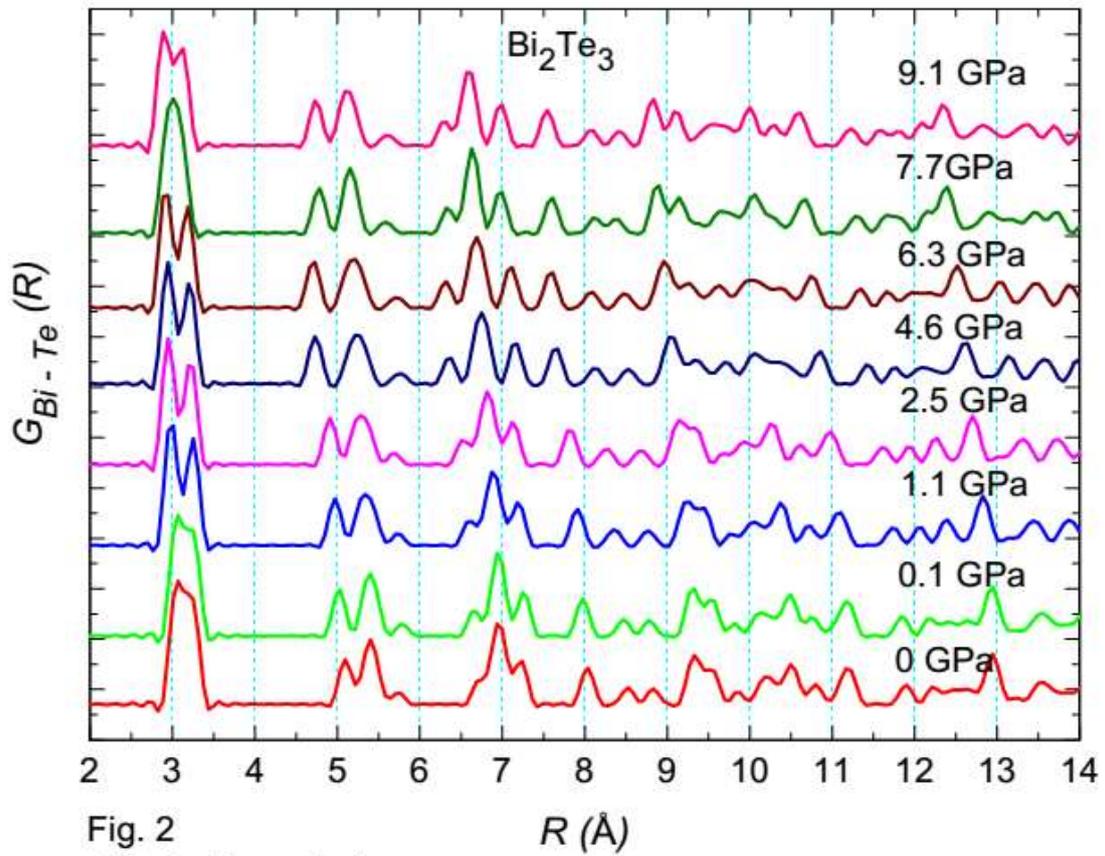

Figure 2: Pressure dependence of $G_{Bi\text{-}Te}(R)$ function for rhombohedral $Bi_2Te_3$.

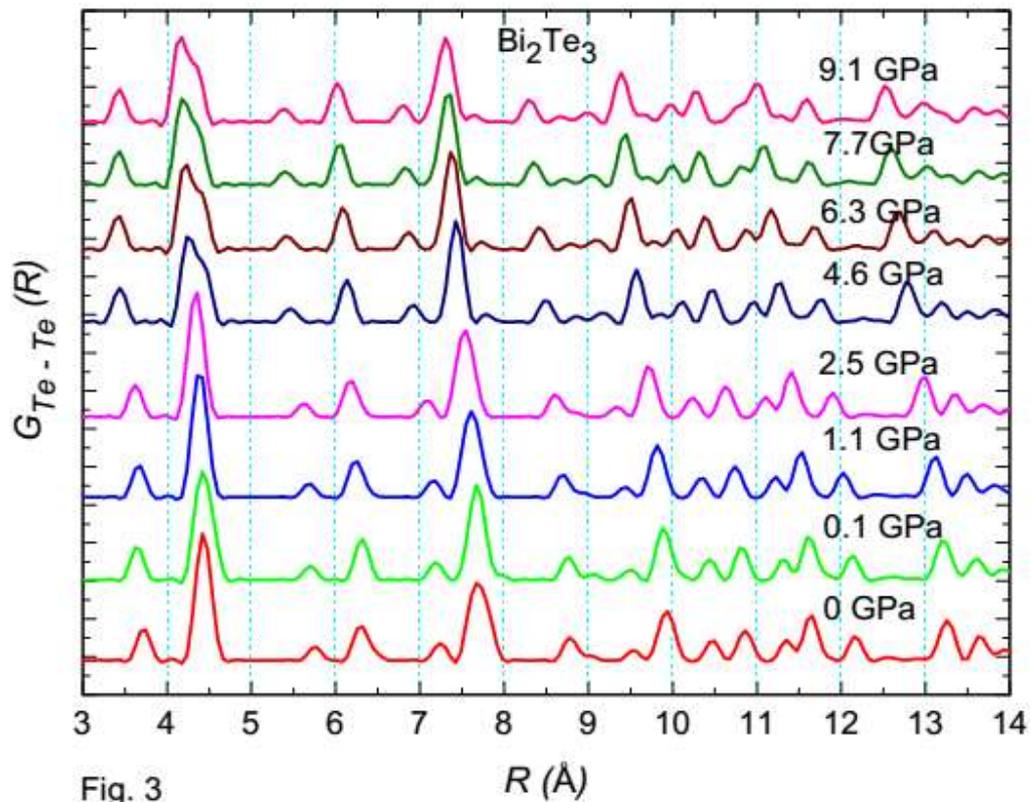

Figure 3: Pressure dependence of $G_{Te\text{-}Te}(R)$ function for rhombohedral $Bi_2Te_3$.

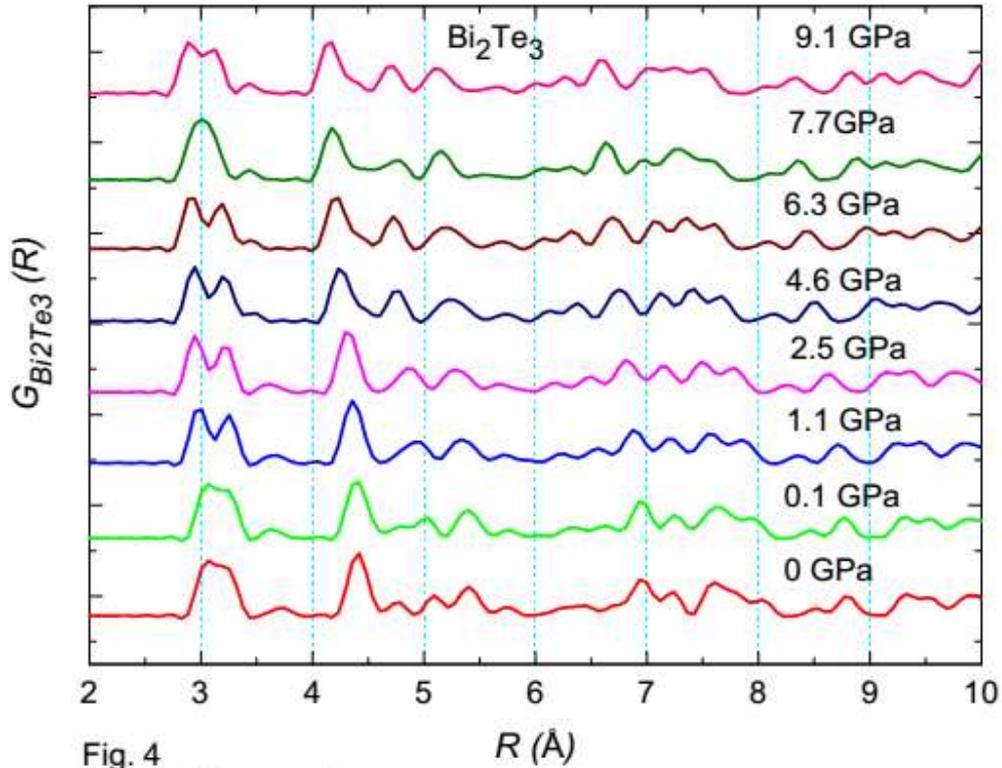

Figure 4: Pressure dependence of $G_{Bi2Te3}(R)$ function for rhombohedral $Bi_2Te_3$.

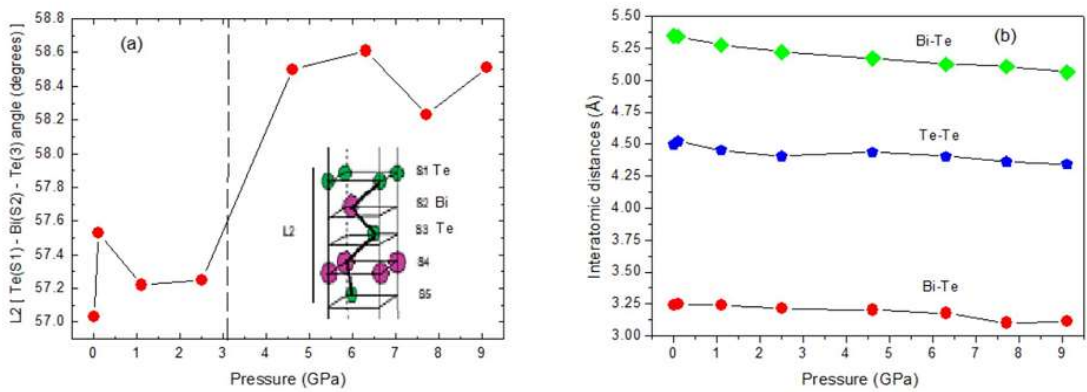

Figure 5: Pressure dependence of intralayer angle L2 [Te(S1)-Bi(S2)-Te(S3)] and intralayer interatomic distances L2 [Te(S1)-Bi(S2), Bi(S2)-Te(S3), and Te(S1)-Te(S3)].

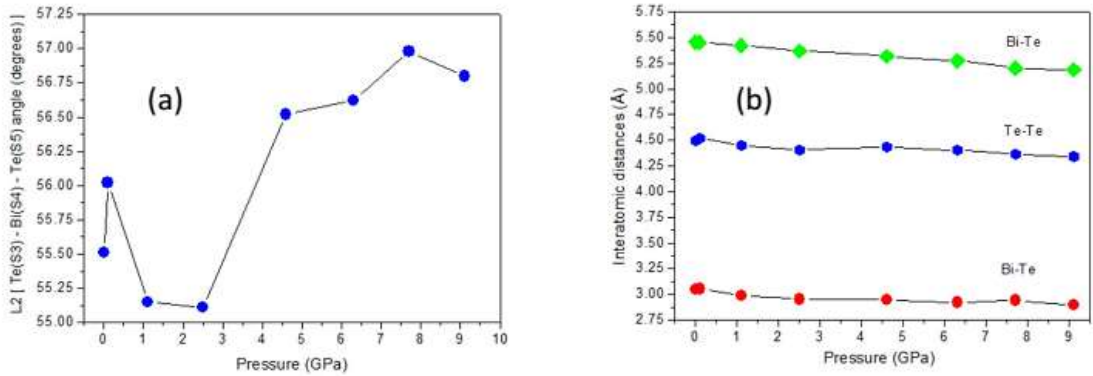

Figure 6: Pressure dependence of intralayer angle L2 [Te(S3)-Bi(S4)-Te(S5)] and intralayer interatomic distances L2 [Te(S3)-Bi(S4), Bi(S4)-Te(S5), and Te(S3)-Te(S5)].

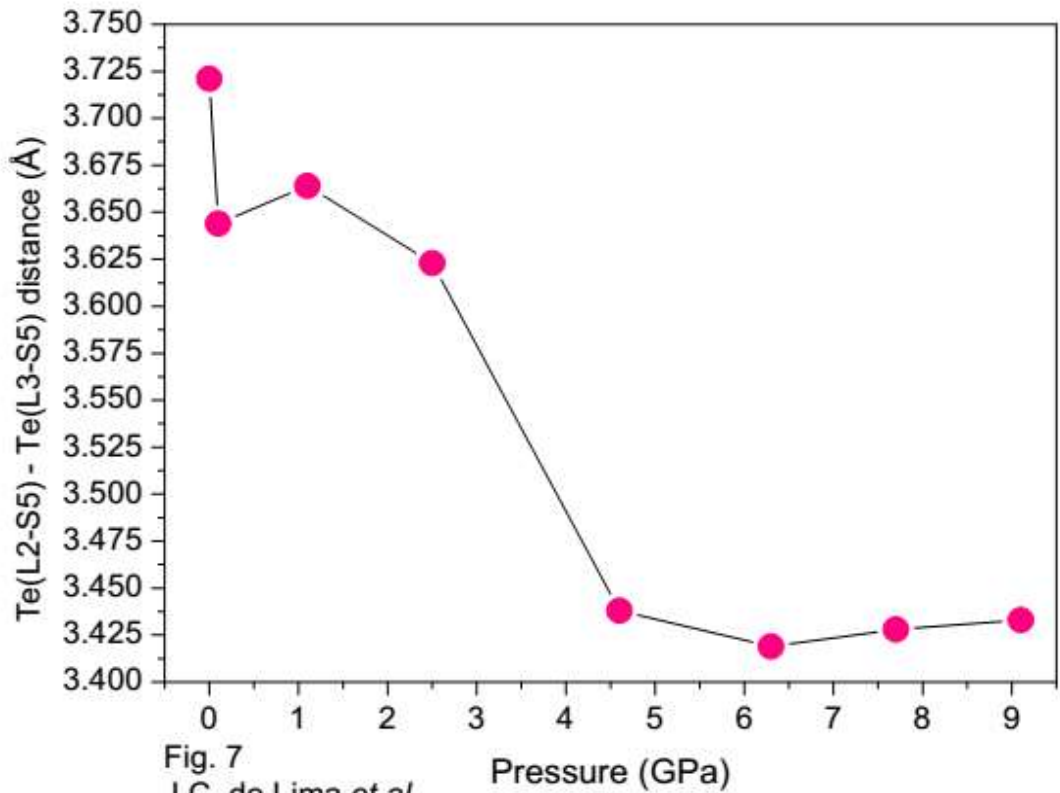

Figure 7: Pressure dependence of interlayers distance L2-Te(S5)-L3-Te(S5) for rhombohedral Bi$_2$Te$_3$.

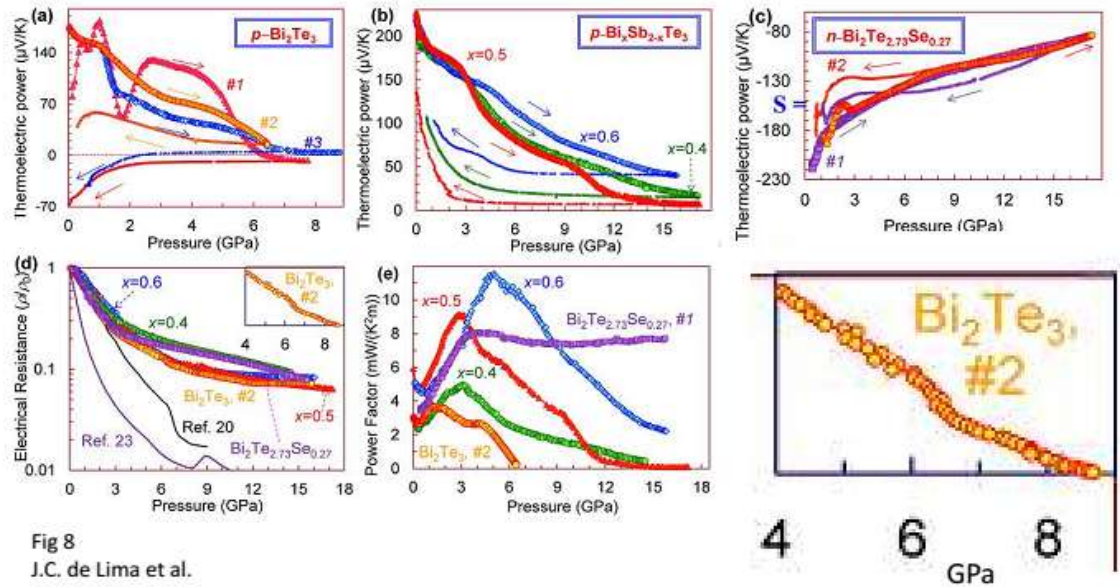

Figure 8: Pressure dependence of macroscopic properties for rhombohedral Bi$_2$Te$_3$. Taken from Ref. [11].

17